\documentclass[a4paper,11pt]{article}

\usepackage{pos}
\usepackage{lipsum} 

\usepackage{caption}
\usepackage{subcaption}
\newlength{\bibitemsep}
\newlength{\bibparskip}
\let\oldthebibliography\thebibliography
\renewcommand\thebibliography[1]{%
  \oldthebibliography{#1}%
  \setlength{\parskip}{\bibitemsep}%
  \setlength{\itemsep}{\bibparskip}%
}

\title{Searching for IceCube sub-TeV neutrino counterparts to sub-threshold Gravitational Wave events}

\ShortTitle{Sub-TeV neutrino counterpart for sub-threshold GW}

\author{The IceCube Collaboration \\{\normalsize \normalfont(a complete list of authors can be found at the end of the proceedings)}\\}

\emailAdd{tista.mukherjee@kit.edu}

\abstract{

Since the release of the Gravitational Wave Transient Catalogue GWTC-2.1 by the LIGO-Virgo collaboration, sub-threshold gravitational wave (GW) candidates are publicly available. They are expected to be released in real-time as well, in the upcoming O4 run. Using these GW candidates for multi-messenger studies complement the ongoing efforts to identify neutrino counterparts to GW events. This in turn, allows us to schedule electromagnetic follow-up searches more efficiently. However, the definition and criteria for sub-threshold candidates are pretty flexible. Finding a multi-messenger counterpart via archival studies for these candidates will help to set up strong bounds on the GW parameters which are useful for defining a GW signal as sub-threshold, thereby increasing their significance for scheduling follow-up searches. Here, we present the current status of this ongoing work with the IceCube Neutrino Observatory. We perform a selection of the sub-threshold GW candidates from GWTC-2.1 and conduct an archival search for sub-TeV neutrino counterparts detected by the dense infill array of the IceCube Neutrino Observatory, known as "DeepCore". For this, an Unbinned Maximum Likelihood (UML) method is used. We report the 90\% C.L. sensitivities of this sub-TeV neutrino dataset for each selected sub-threshold GW candidate, considering the spatial and temporal correlation between the GW and neutrino events within a 1000~s time window.

\vspace{4mm}
{\bfseries Corresponding author:}
Tista Mukherjee$^{1*}$ 
\\
{$^{1}$ \itshape Institute for Astroparticle Physics, Karlsruhe Institute of Technology, Karlsruhe, Germany}\\
$^*$ Presenter

\ConferenceLogo{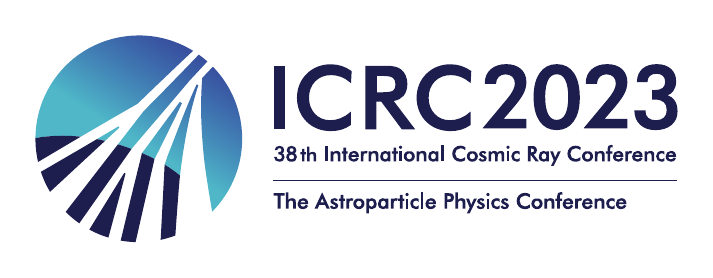}
\FullConference{The 38th International Cosmic Ray Conference (ICRC2023)\\ 26 July -- 3 August, 2023\\ Nagoya, Japan}
}
\begin{document}

\maketitle

\section{Introduction}\label{sec1}
The IceCube Neutrino Observatory is a cubic-kilometre ice-Cherenkov detector located in the geographical South Pole. It is mainly sensitive to high-energy neutrinos with an energy of more than 100 GeV. However, after the installation of a dense infill array called `DeepCore', the energy threshold for neutrino detection by IceCube has been lowered by another order of magnitude, making the detector sensitive to sub-TeV neutrinos with energy < 100 GeV as well \cite{deepcore}.

IceCube, with all its resources, has been playing an active role, both in real-time and archival follow-ups to the gravitational wave (GW) alerts sent by LIGO-Virgo-KAGRA (LVK) collaboration \cite{GWfollowupO3, gw_nu}. These alerts were sent for the confident events observed during their past three observation runs (O1, O2 and O3). By confident events, we mean events with False Alarm Rate (FAR) < 2 $\mathrm{yr^{-1}}$ and the probability of astrophysical origin ($p_\mathrm{astro}) > 0.5$.
The events are currently listed in the three event catalogues released by LVK, known as Gravitational Wave Transient Catalogue (GWTC) - 1, 2 and 3 \cite{gwtc-1, gwtc-2, gwtc-3}.
However, after the release of GWTC-2, GWTC-2.1 was made public with an updated list of confident GW events from the first half of third observational run (O3a) \cite{gwtc-2.1}. Another feature which made the GWTC-2.1 unique was the inclusion of 1201 `sub-threshold candidates' detected during O3a. By `sub-threshold', we mean GW candidates which could not pass the hard threshold of FAR < 2 $\mathrm{yr^{-1}}$, but they passed a low-significance FAR threshold of 2 $\mathrm{day^{-1}}$. Similarly, 1048 sub-threshold events were further added to the GWTC-3 catalogue as well, detected during the second half of the third observational run (O3b).

From IceCube follow-ups to the confident GW events, no neutrino counterpart was detected \cite{GWfollowupO3, greco-gw}. However, the sub-threshold GW candidates were only available after the end of O3. So, they can only be followed up via archival studies. In this work, we are interested in doing that using the candidates we are getting from O3a. The main motivation to search for possible neutrino counterparts of sub-threshold GW candidates is to check if they are significant enough to be followed up in real-time. Also, with the help of IceCube, we can significantly improve the source localisation for sub-threshold GW candidates, which would allow the astronomical community to schedule follow-up campaigns more efficiently.

\section{The datasets}
\subsection{GW sub-threshold candidates}\label{sec2}
\begin{figure}[ht!]
	\centering
            \includegraphics[width=11cm]{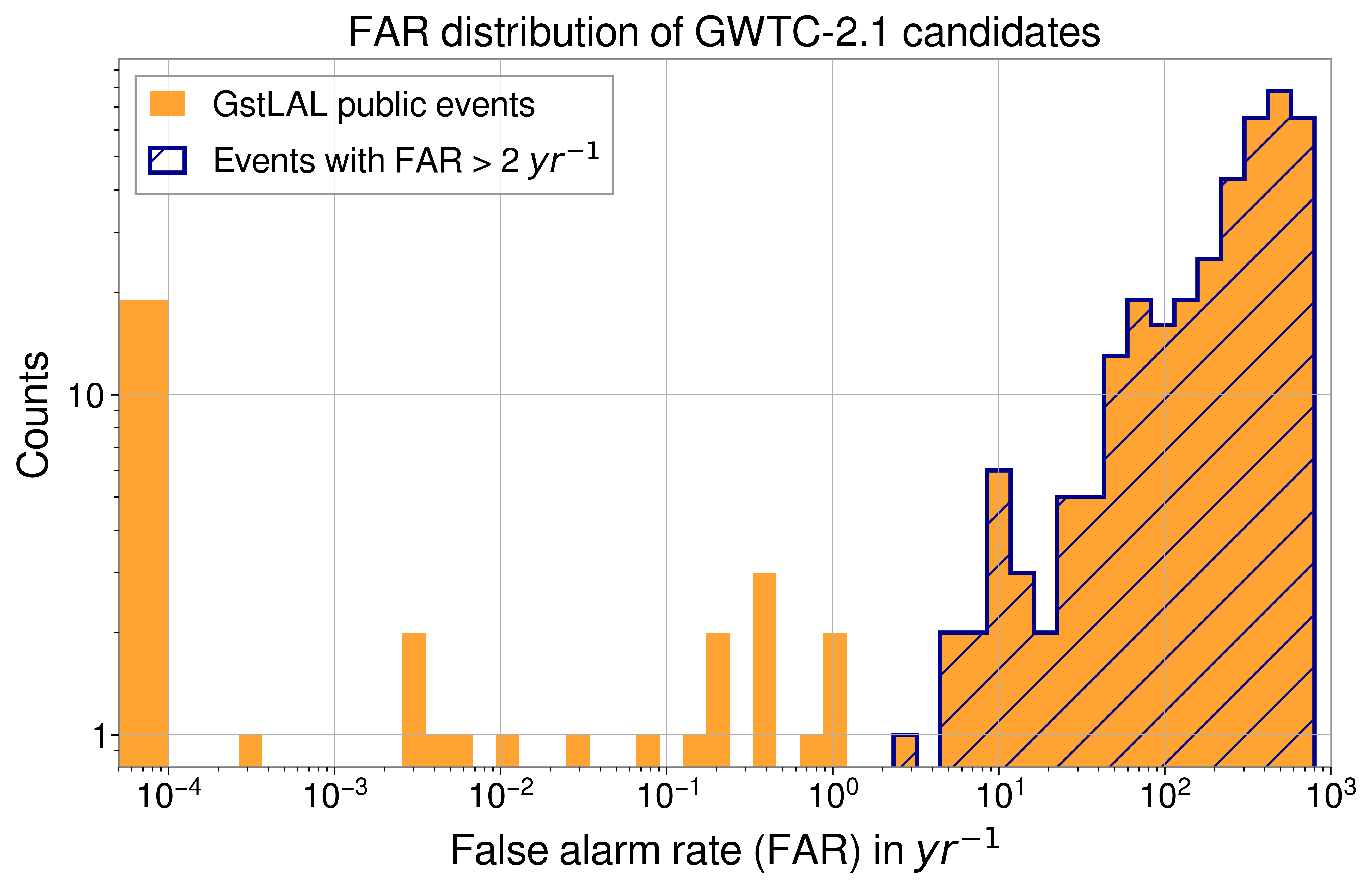}   
	\caption{The comparison of the FAR distribution with and without the confident GW events identified by GstLAL during O3a. It is to be noted that the FAR values range from ~ $10^{-33}$ to ~730 $\mathrm{yr^{-1}}$ (~0 - 2 $\mathrm{day^{-1}}$). The first bin contains all candidates with FAR < $10^{-4}$ $\mathrm{yr^{-1}}$. After sorting out the confident events, we lose the GW candidates identified by GstLAL with FAR < 2 $\mathrm{yr^{-1}}$. The shaded region represent all sub-threshold candidates with FAR > 2 $\mathrm{yr^{-1}}$ from GstLAL. Similar treatment is done for events identified by other three pipelines as well.}
        \label{Fig:GstLAL_subth}
\end{figure}
All the GW candidates are identified using four different GW analysis pipelines, i.e., GstLAL, MBTA, PyCBC, and PyCBC-high mass \cite{gstlal, mbta, pycbc, pycbc-HM}, developed and used by the LVK collaboration to detect compact binary coalescence (CBC). In the event database for GWTC-2.1, all candidates with FAR < 2 $\mathrm{day^{-1}}$ ($\sim$ 730 $\mathrm{yr^{-1}}$) were included. This also indicates that the confident events with FAR < 2 $\mathrm{yr^{-1}}$ are not separated from the candidates of our interest for this work. So first, we need to filter out these candidates qualifying as confident GW events for which archival searches have already been done using both high- and low-energy IceCube neutrinos \cite{greco-gw, GWfollowupO3}. In \autoref{Fig:GstLAL_subth}, we have  illustrated the procedure, taking the candidates we are getting from GstLAL as an example. Initially, it included 405 candidates with FAR values ranging from $5 \times 10^{-33}$ to $\sim$ 730 $\mathrm{yr^{-1}}$. After excluding all the events with FAR < 2 $\mathrm{yr^{-1}}$, we are left with 369 sub-threshold candidates. Similar treatment is done to the candidates identified by the other 3 pipelines.
\begin{figure}[ht!]
	\centering
            \includegraphics[width=11cm]{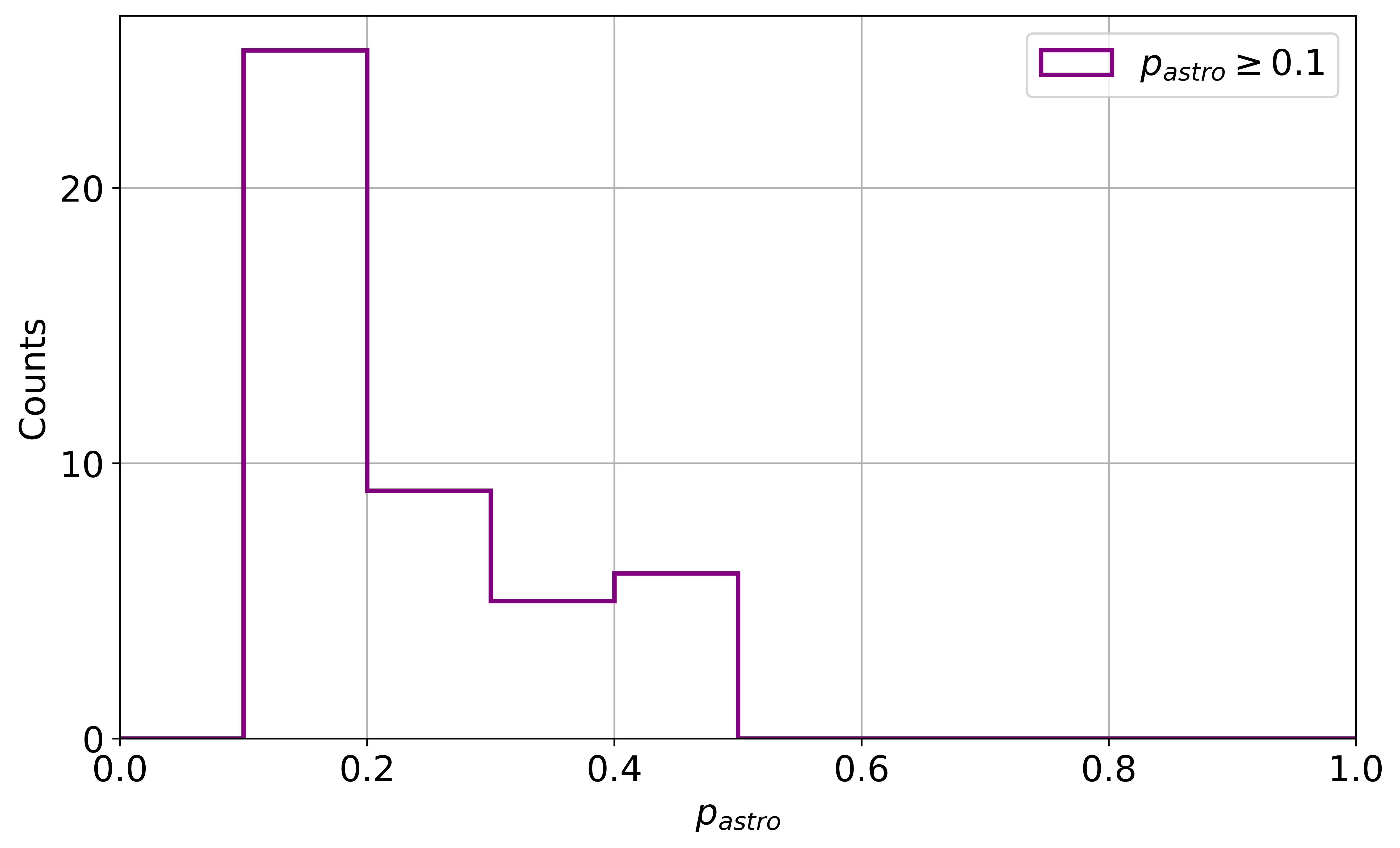}
	\caption{$p_\mathrm {astro}$ distribution for the 45 sub-threshold GW candidates detected during O3a, to be considered for neutrino counterpart search.}
	\label{Sub-th_pastro_selected}
\end{figure}

Some of these sub-threshold candidates were later qualified as confident events in the updated GWTC-2.1 catalogue as they had higher $p_\mathrm{astro}$ or higher SNR or both. We also remove those candidates from our list as they were previously included in the list of GW candidates for archival searches by IceCube. On the leftover candidates, we further impose another filtering criterion of eliminating all the sub-threshold candidates with $p_\mathrm{astro} < 0.1$ as they have a higher probability of being terrestrial noise. Some candidates were also double counted by multiple pipelines. In this case, we only consider the candidate with the highest $p_\mathrm{astro}$ value from one particular pipeline, at a given event-time. In the end, we are left with 45 unique sub-threshold events which are not included in the updated GWTC-2.1 confident event list. The $p_\mathrm{astro}$ distribution for these candidates is shown in \autoref{Sub-th_pastro_selected}.

\subsection{The sub-TeV neutrino dataset (GRECO)}
\begin{figure}[t]
	\centering
	\includegraphics[width=10cm]{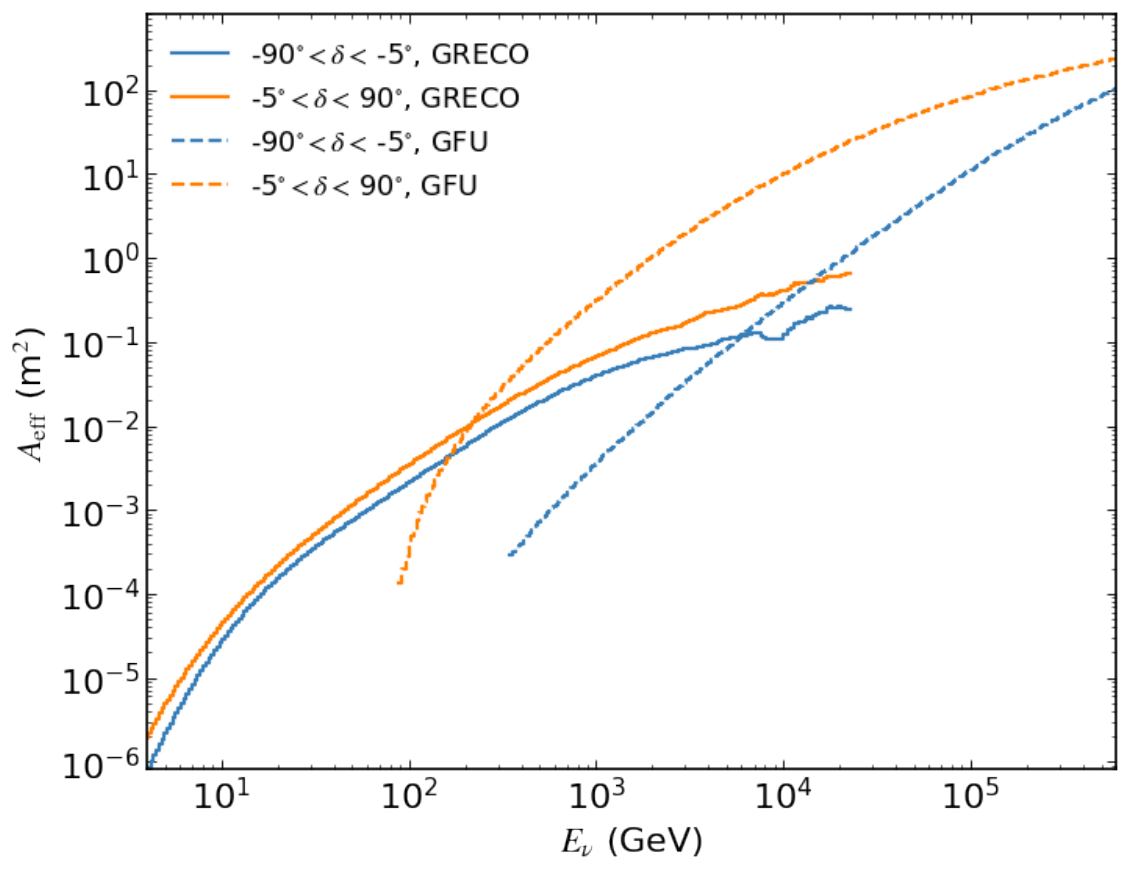}
	\caption{All-flavour effective area for GRECO in different declination bands from \cite{novae}. The GFU effective area is shown (dashed line) as reference to compare. It is evident that GRECO has comparable effective area coverage all over the sky, that compliments GFU (sensitive to ~O(10-100)TeV energy) well in the lower energy range.}
	\label{Aeff}
\end{figure}

The sub-TeV neutrino dataset available within IceCube is called `GRECO' (GeV Reconstructed Events with Containment for Oscillations). This is an all-sky, all-flavour dataset, covering the neutrino energy range of O(10-100) GeV. The GRECO events were originally used for neutrino oscillation studies. However, it also has a very stable event rate which makes it suitable for transient searches. It also has a very good effective-area coverage that energy range, evident from \autoref{Aeff}.
It complements the effective area of higher-energy Gamma-ray Follow Up (GFU) events in the energy range of O(10-100) TeV, which is a standard dataset often used to search for astrophysical neutrino transients. For these reasons, the dataset was further optimized for probing transient astrophysical sources from which we can expect sub-TeV neutrinos \cite{greco}. Previously, we have used GRECO to search for low-energy neutrino counterparts from the 90 confident GW events from O1, O2 and O3 \cite{greco-gw}, also from novae \cite{novae} and GRBs \cite{greco_grb}. In this analysis, it is considered for low-energy neutrino counterpart search to the selected sub-threshold GW candidates from O3a.

\section{Analysis Method}
To search for the neutrino counterparts, we will follow a Unbinned Maximum Likelihood (UML) analysis. We look for a spatial and temporal coincidence between the selected sub-threshold GW candidates and GRECO events within a 1000 s ($\pm 500$ s) time window around the GW event-time. So, in this case, we define two competing hypotheses:
\\
Null hypothesis ($H_\mathrm{0}$): We only have background neutrino events in our data.
\\
Signal hypothesis ($H_\mathrm{s}$): We have injected some signal neutrinos correlated with a specific GW candidate, along with background events.

To test our hypotheses, we define a likelihood function $\mathscr{L}$ ($n_\mathrm{s}, \gamma$), which is given as
\begin{equation}
\label{likelihood}
    \mathscr{L} (n_\mathrm{s} (\gamma)) = \frac{(n_\mathrm{s} + n_\mathrm{b})^N}{N!} \mathrm e^{-(n_\mathrm{s} + n_\mathrm{b})} \prod_{i = 1}^{N} \Big ( \frac{n_\mathrm{s} S_\mathrm{i}}{n_\mathrm{s} + n_\mathrm{b}} + \frac{n_\mathrm{b} B_\mathrm{i}}{n_\mathrm{s} + n_\mathrm{b}} \Big ).
\end{equation}
Here, $n_\mathrm{s}$ is and $n_\mathrm{b}$ are signal and background neutrino events respectively, where $n_\mathrm{s} + n_\mathrm{b} = N$, which is the total number of neutrinos. For the signal neutrinos, $\gamma$ is the spectral index. It is not explicitly shown in the likelihood function `$\mathscr{L}$' because it is implicit in $n_\mathrm{s}$. A Poisson term $\mathrm{e}^{-(n_\mathrm{s} + n_\mathrm{b})}$ arises to describe the transient nature of the source. $S_\mathrm{i}$ and $B_\mathrm{i}$ are signal and background probability density for $i^{th}$ event.

We can calculate the likelihood for each hypothesis. Then, we test our $H_\mathrm{s}$ where we have some signal neutrinos, $n_\mathrm{s}$ which follow a spectrum with spectral index $\gamma$, with respect to the background-like $H_\mathrm{0}$. We define the `Test Statistic' (TS) as
\begin{equation}
\label{TS}
    TS = max. \Bigg[ 2~\mathrm{ln}  \Bigg( \frac{\mathscr{L}_\mathrm{k} (n_\mathrm{s}(\gamma)) \cdot \omega_\mathrm{k}}{\mathscr{L}_\mathrm{k} (n_\mathrm{s} = 0) } \Bigg) \Bigg ]
    = TS_{PS} + 2 \ln (\omega_\mathrm{k}).
\end{equation}
As here we are dealing with GW skymaps to search for spatial correlation, we have a spatial prior term $\omega_{k}$ in the numerator of TS, where $k$ is the ID of each pixel in the sky. We divide the GW skymap into 49152 equally-sized pixels. For each pixel, we calculate $\omega_\mathrm{k}$ as $\frac{(P_{GW})_\mathrm{k}}{\Omega_\mathrm{pixel}}$ where $(P_{GW})_\mathrm{k}$ is the probability of having a GW source in the $k^{th}$ pixel. Now for each pixel,
\begin{equation}
\label{prior}
    2 \ln (\omega_\mathrm{k}) = [2 \ln (\omega_\mathrm{k})] - max. [2 \ln (\omega_\mathrm{k})].
\end{equation}
So, the maximum value of 2 ln ($\omega_\mathrm{k}$) is 0 for the best-fit GW source location. It is increasingly negative for the pixels where the GW source is less and less likely to be located. Thus, it acts as a spatial penalty term depending on the GW skymap.

\section{Analysis Performance}
Before starting the analysis, we need to construct the background TS (BGTS) distribution for each sub-threshold GW candidate. This is done by randomly sampling the arrival times of each neutrino event which was detected $\pm$ 5 days around the time the GW signal was recorded. This task is repeated 10,000 times, which are designated as individual `pseudo experiments'. This BGTS distribution is then compared against the TS values we get from an unscrambled neutrino dataset, to search for correlated signal neutrino events. But before doing that, we test the strength of our analysis by `injecting' signal-like neutrino events from a Monte-Carlo dataset with $\gamma = 2$ on the background-like, scrambled dataset. From this we calculate the 90\% sensitivity. This is reported for each selected sub-threshold GW candidate.

\subsection{Background TS distribution}
\begin{figure}
    \centering{}
    \begin{subfigure}[h]{0.49\textwidth}
         \centering
         \includegraphics[width=\textwidth]{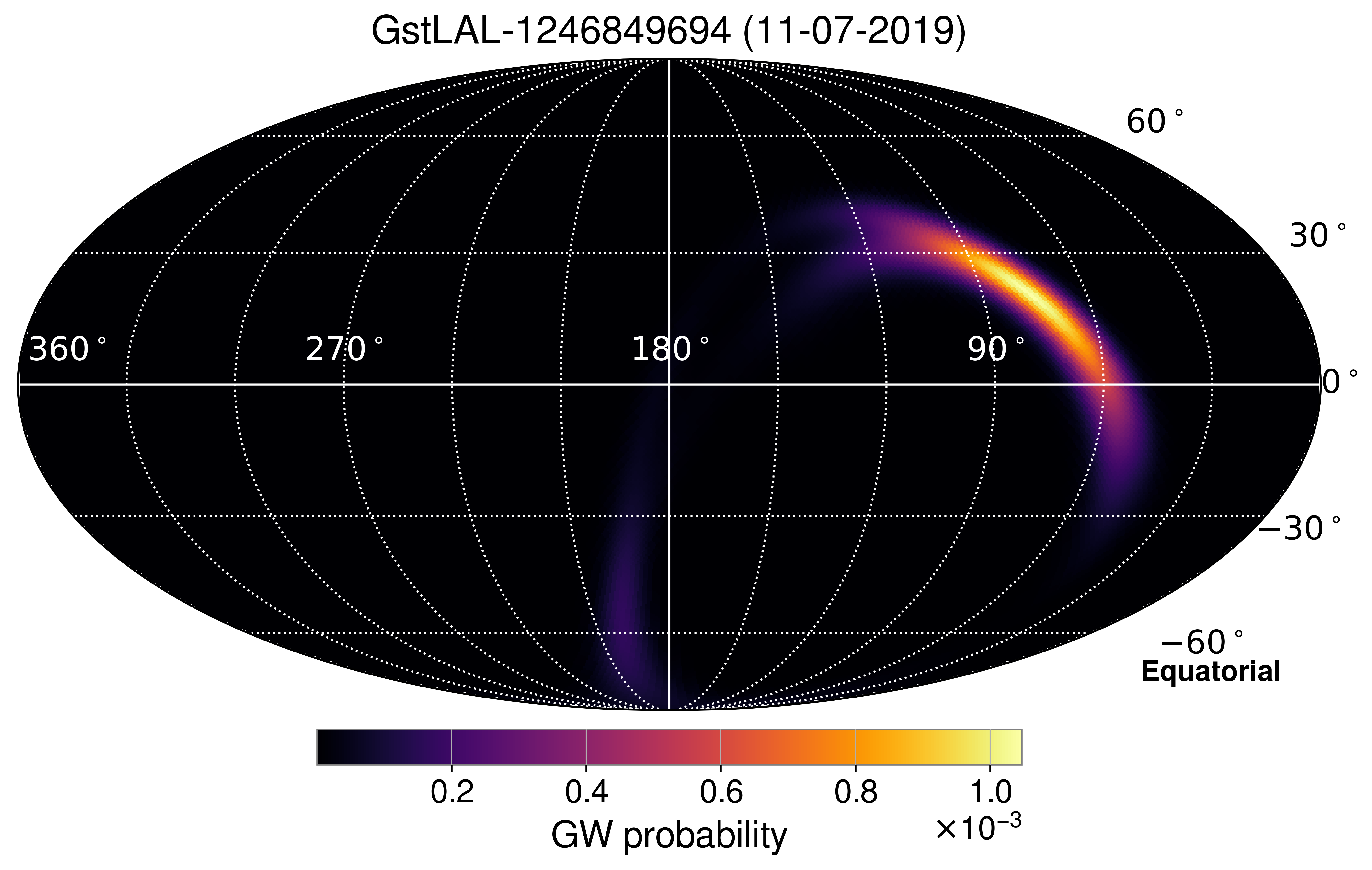}
         \caption{}
     \end{subfigure}
     \hfill
     \begin{subfigure}[h]{0.5\textwidth}
         \centering
         \includegraphics[width=\textwidth]{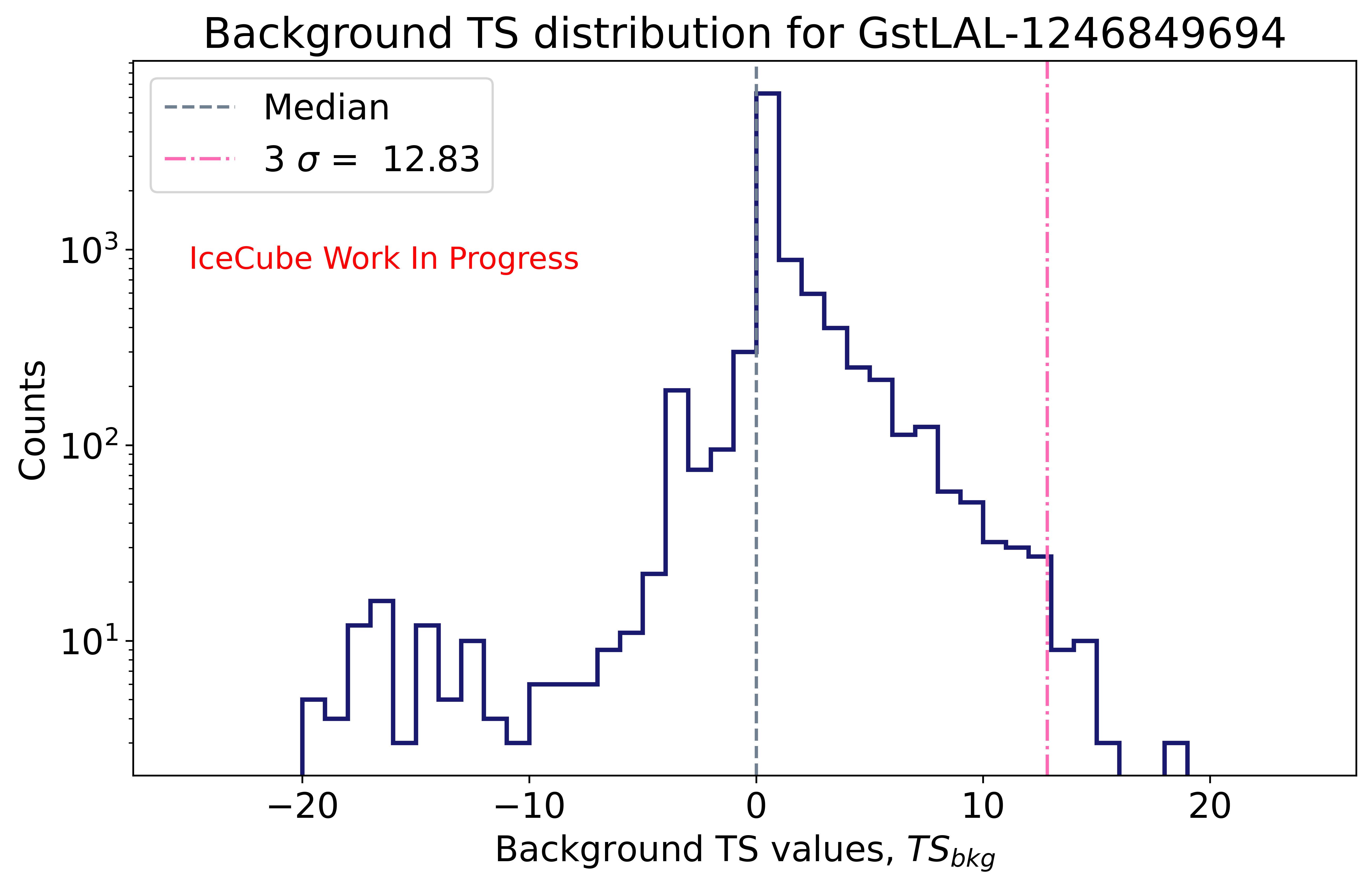}
         \caption{}
     \end{subfigure}
	\caption{Here we consider the sub-threshold GW candidate GstLAL-1246849694 ($p_\mathrm{astro}$ = 0.35), detected during O3a, as an example. We present (a) the corresponding skymap for this candidate, and (b) the background TS distribution including spatial priors. It contains several negative TS values for the background neutrinos falling on the pixels with highly negative spatial penalty term.}
        \label{BkgTS}
\end{figure}
In \autoref{BkgTS}, we present the BGTS distribution we get from one of the sub-threshold GW candidates (GstLAL-1246849694) detected by the LIGO-Virgo-KAGRA collaboration on July 11, 2019, during O3a. The corresponding GW skymap is also shown. One can see that the BGTS distribution contains some dominant spatial features which it gets from $\omega_\mathrm{k}$. Most of the BGTS values are negative as most of the background neutrinos fall on the pixels where 2 ln($\omega_\mathrm{k}$) is highly negative.

\subsection{90\% Sensitivity flux}
\begin{figure}[b]
    \centering
    \begin{minipage}{0.75\textwidth}
        \centering
        \includegraphics[width = \linewidth]{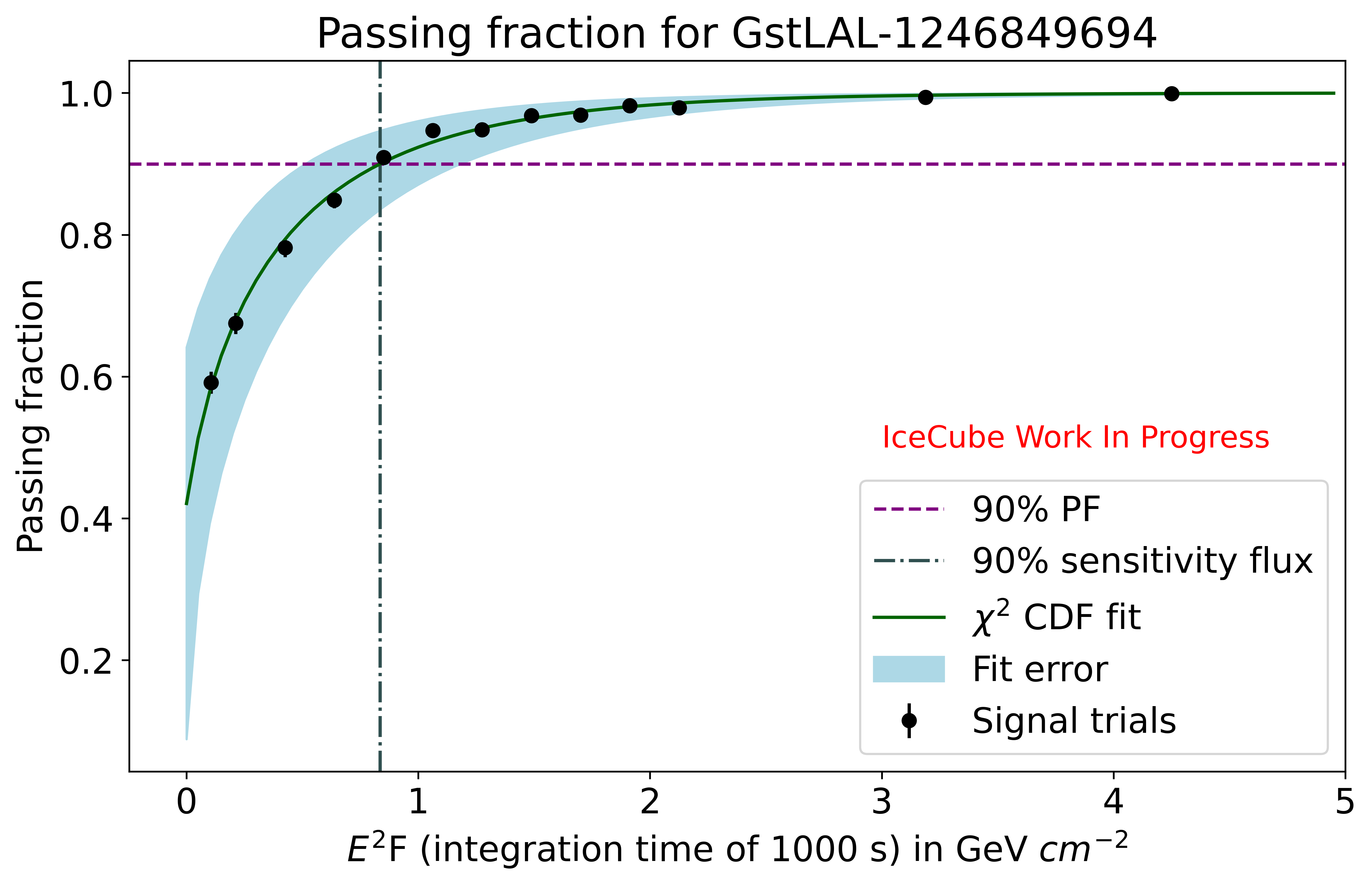}
    \end{minipage}
    \begin{minipage}{0.24\textwidth}
        \caption{Passing fraction curve for GstLAL-1246849694. The 90\% sensitivity flux is pointed out. The time-integrated flux, F (in $\mathrm{GeV^{-1}cm^{-2}}$) = $\frac{dN}{dEdAdt} \cdot \delta t$ is computed for $\delta t =$ 1000 s time window.}
        \label{PassingFraction}
        \end{minipage}
\end{figure}
The 90\% sensitivity is defined as the flux for which the TS value after signal injection is greater than the median of the BGTS distribution for 90\% of the trials. We calculate this by injecting different $n_s$ values on the GW skymap to generate different TS values, and compare that with the BGTS distribution. This is also repeated for 10,000 times for each sub-threshold GW candidate. For each different $n_s$ value, we will have a different flux recorded by our detector. We calculate for each flux level, the Passing Fraction (PF), which is the fraction of the pseudo experiments where, for neutrino injection, we have a TS value higher than the median of BGTS. The flux for which we obtain PF~=~0.9, is our 90\% sensitivity flux for the given dataset. In \autoref{PassingFraction}, we show the PF curve for S1246849694 as an example.

\begin{figure}[t]
     \begin{subfigure}{0.49\textwidth}
         \centering
         \includegraphics[width=\textwidth]{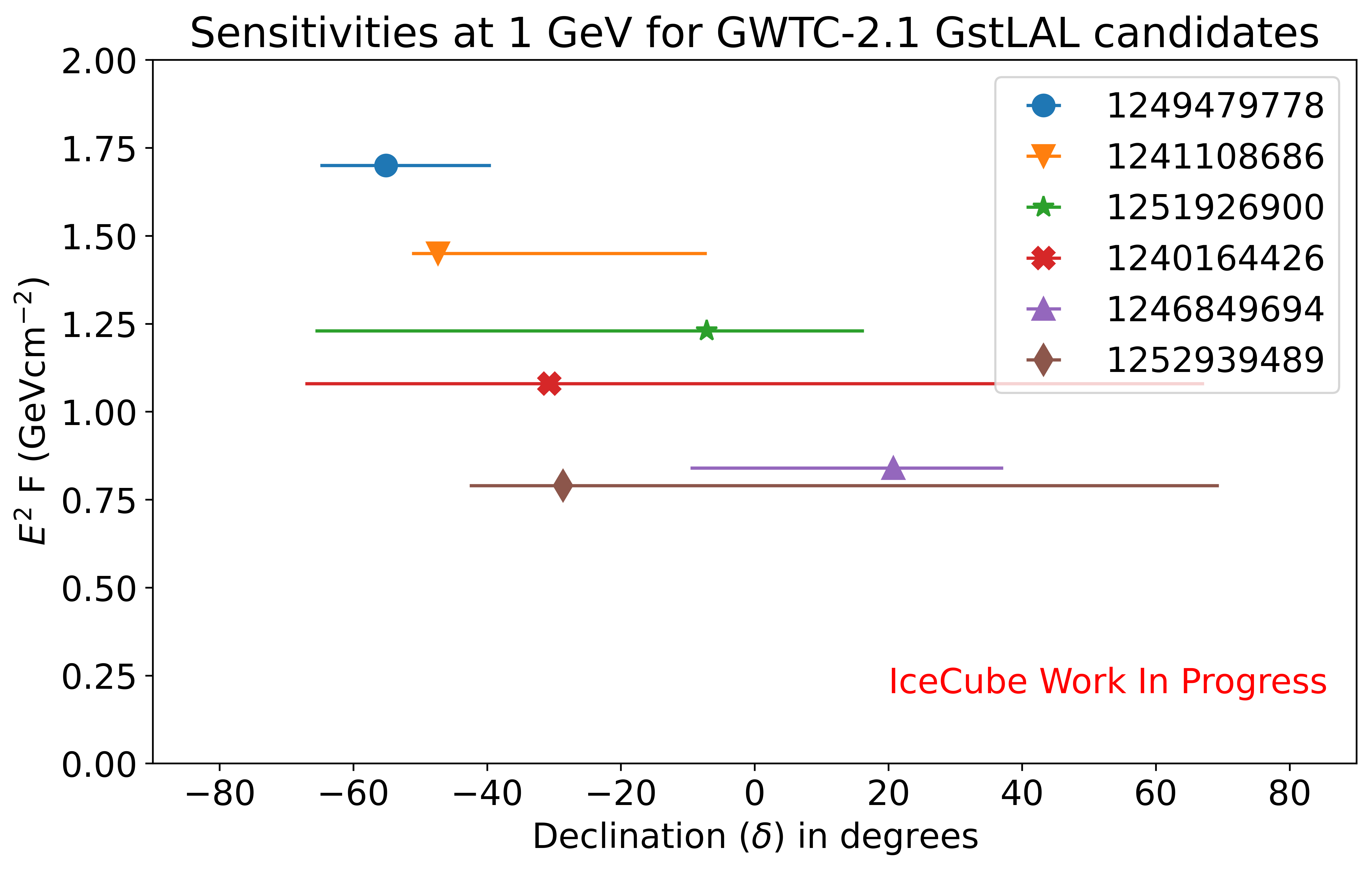}
         \caption{}
     \end{subfigure}
     \hfill
     \begin{subfigure}{0.49\textwidth}
         \centering
         \includegraphics[width=\textwidth]{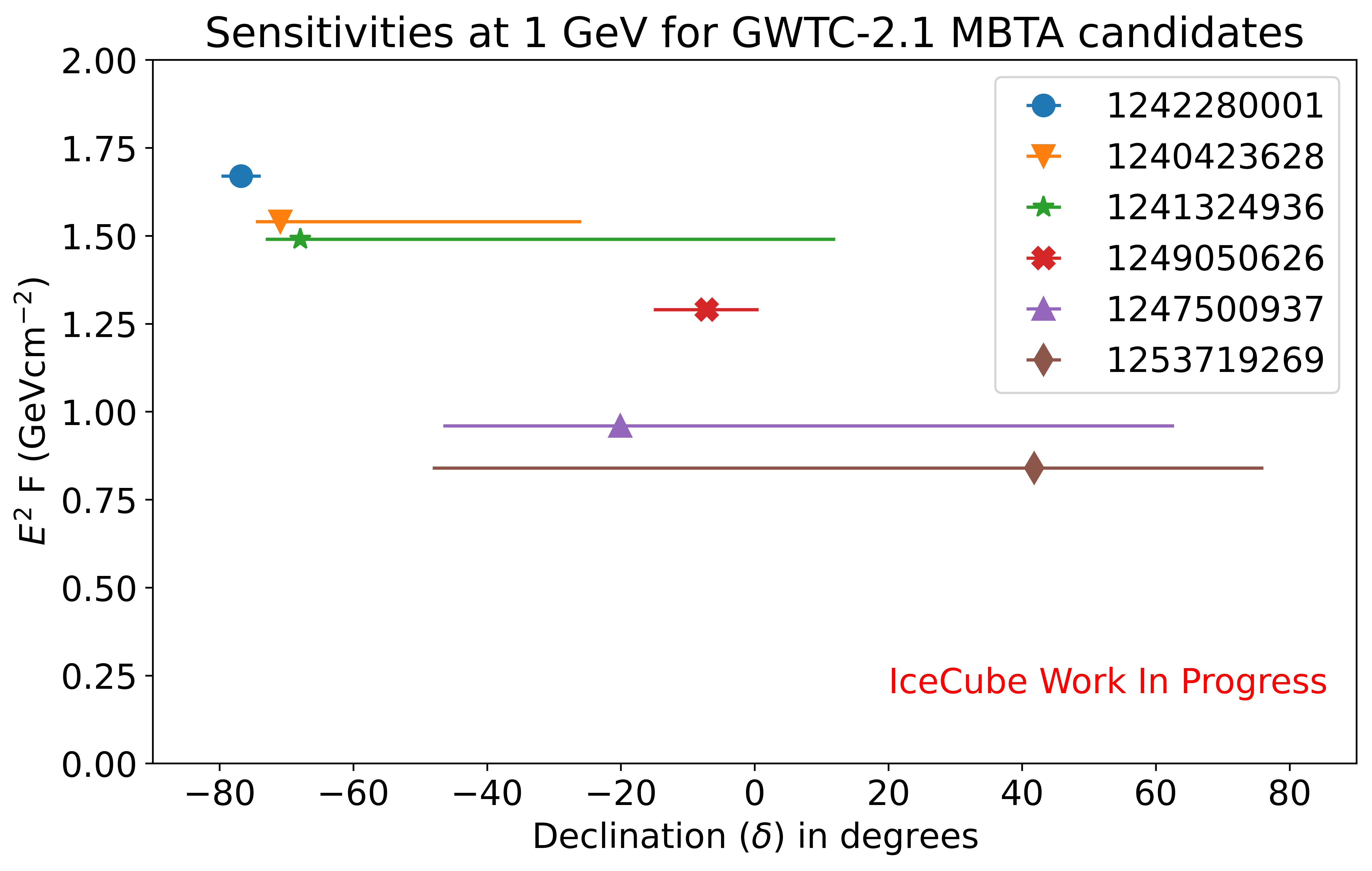}
         \caption{}
     \end{subfigure}
     \hfill
     \begin{subfigure}{0.95\textwidth}
         \centering
         \includegraphics[width=\textwidth]{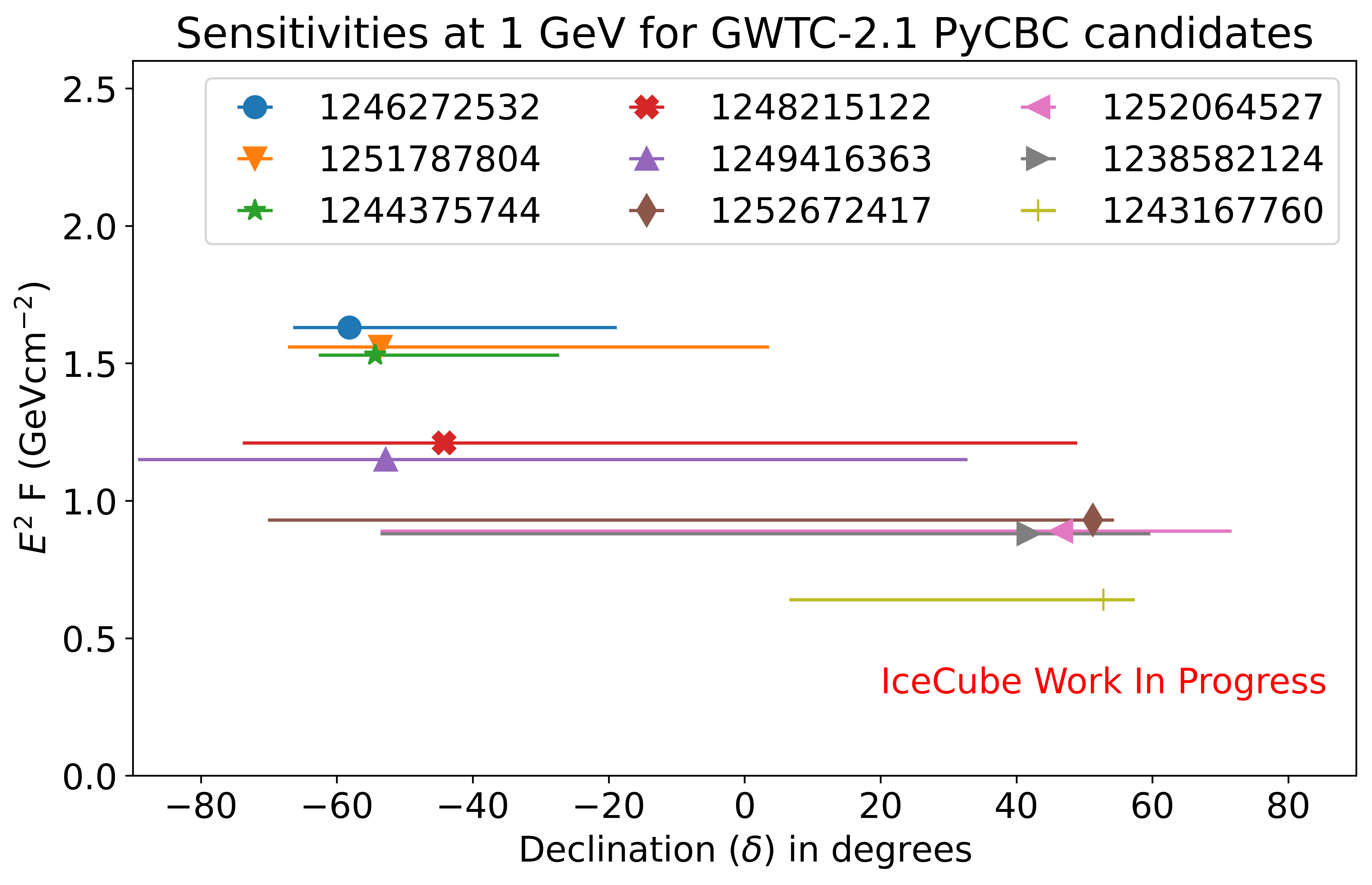}
         \caption{}
     \end{subfigure}
         \caption{Declination dependence of per-flavour sensitivity flux calculated at 1 GeV for the sub-threshold candidates identified by (a) GstLAL (b) MBTA (c) PyCBC during O3a. The declination of the pixel containing with the best-fit source location has been shown by a marker.
        The horizontal spread across the markers covers the declination range for 50\% containment area from the respective GW skymap.}
    \label{Sens_vs_Dec}
\end{figure}

We also study the declination dependence of per-flavour sensitivity of the sub-threshold GW candidates. In \autoref{Sens_vs_Dec}, we show the sensitivities of the sub-threshold candidates identified by the GstLAL, MBTA and PyCBC pipeline. From this figure, we see the general behaviour is to have better sensitivities in the Northern Hemisphere than in the Southern Hemisphere. This is a well-known feature of GRECO which was also identified in a previous analysis \cite{greco-gw}. Also, it is notable that even the 50\% containment region for some of the sub-threshold candidates is pretty large, ranging from the Northern to the Southern hemisphere. The main reason for this is worse localisation of sub-threshold candidates by the GW detectors. However, this could be significantly improved if we can identify neutrino counterparts spatially and temporally correlated to these candidates.
\section{Outlook}
Neutrino counterpart search for confident GW events have been conducted by IceCube over all possible energy ranges, from few GeV to O(100) GeV. In this work, we look for spatially coincident sub-TeV neutrinos with sub-threshold GW candidates, within 1000 s time window. We have completed per-flavour sensitivity studies for the GW candidates with $p_\mathrm{astro}$ > 0.1. Declination dependence of per-flavour sensitivity has been shown for candidates from three different pipelines - GstLAL, MBTA and PyCBC. Notably, for some sub-threshold GW candidates, the localisation is particularly worse. If identifying coincident neutrino events is possible, the GW source localisation could be significantly improved. This would not only make future follow-ups of sub-threshold GW candidates more efficient, but also help us to gather maximum information from the GW progenitors.

\bibliographystyle{ICRC}
\bibliography{references}
%
\clearpage
\section*{Full Author List: IceCube Collaboration}

\scriptsize
\noindent
R. Abbasi$^{17}$,
M. Ackermann$^{63}$,
J. Adams$^{18}$,
S. K. Agarwalla$^{40,\: 64}$,
J. A. Aguilar$^{12}$,
M. Ahlers$^{22}$,
J.M. Alameddine$^{23}$,
N. M. Amin$^{44}$,
K. Andeen$^{42}$,
G. Anton$^{26}$
% ...
\\
\\
% ...
$^{12}$ Universit{\'e} Libre de Bruxelles, Science Faculty CP230, B-1050 Brussels, Belgium \\
% ...
$^{17}$ Department of Physics, Loyola University Chicago, Chicago, IL 60660, USA \\
$^{18}$ Dept. of Physics and Astronomy, University of Canterbury, Private Bag 4800, Christchurch, New Zealand \\
% ...
$^{22}$ Niels Bohr Institute, University of Copenhagen, DK-2100 Copenhagen, Denmark \\
$^{23}$ Dept. of Physics, TU Dortmund University, D-44221 Dortmund, Germany \\
% ...
$^{26}$ Erlangen Centre for Astroparticle Physics, Friedrich-Alexander-Universit{\"a}t Erlangen-N{\"u}rnberg, D-91058 Erlangen, Germany \\
% ...
$^{40}$ Dept. of Physics and Wisconsin IceCube Particle Astrophysics Center, University of Wisconsin{\textendash}Madison, Madison, WI 53706, USA \\
% ...
$^{42}$ Department of Physics, Marquette University, Milwaukee, WI, 53201, USA \\
% ...
$^{44}$ Bartol Research Institute and Dept. of Physics and Astronomy, University of Delaware, Newark, DE 19716, USA \\
% ...
$^{63}$ Deutsches Elektronen-Synchrotron DESY, Platanenallee 6, 15738 Zeuthen, Germany  \\
$^{64}$ Institute of Physics, Sachivalaya Marg, Sainik School Post, Bhubaneswar 751005, India \\
% ...

\subsection*{Acknowledgements}

\noindent
The authors gratefully acknowledge the support from the following agencies and institutions:
USA {\textendash} U.S. National Science Foundation-Office of Polar Programs,
U.S. National Science Foundation-Physics Division,
U.S. National Science Foundation-EPSCoR,
Wisconsin Alumni Research Foundation,
Center for High Throughput Computing (CHTC) at the University of Wisconsin{\textendash}Madison,
Open Science Grid (OSG),
Advanced Cyberinfrastructure Coordination Ecosystem: Services {\&} Support (ACCESS),
Frontera computing project at the Texas Advanced Computing Center,
U.S. Department of Energy-National Energy Research Scientific Computing Center,
Particle astrophysics research computing center at the University of Maryland,
Institute for Cyber-Enabled Research at Michigan State University,
and Astroparticle physics computational facility at Marquette University;
Belgium {\textendash} Funds for Scientific Research (FRS-FNRS and FWO),
FWO Odysseus and Big Science programmes,
and Belgian Federal Science Policy Office (Belspo);
Germany {\textendash} Bundesministerium f{\"u}r Bildung und Forschung (BMBF),
Deutsche Forschungsgemeinschaft (DFG),
Helmholtz Alliance for Astroparticle Physics (HAP),
Initiative and Networking Fund of the Helmholtz Association,
Deutsches Elektronen Synchrotron (DESY),
and High Performance Computing cluster of the RWTH Aachen;
Sweden {\textendash} Swedish Research Council,
Swedish Polar Research Secretariat,
Swedish National Infrastructure for Computing (SNIC),
and Knut and Alice Wallenberg Foundation;
European Union {\textendash} EGI Advanced Computing for research;
Australia {\textendash} Australian Research Council;
Canada {\textendash} Natural Sciences and Engineering Research Council of Canada,
Calcul Qu{\'e}bec, Compute Ontario, Canada Foundation for Innovation, WestGrid, and Compute Canada;
Denmark {\textendash} Villum Fonden, Carlsberg Foundation, and European Commission;
New Zealand {\textendash} Marsden Fund;
Japan {\textendash} Japan Society for Promotion of Science (JSPS)
and Institute for Global Prominent Research (IGPR) of Chiba University;
Korea {\textendash} National Research Foundation of Korea (NRF);
Switzerland {\textendash} Swiss National Science Foundation (SNSF);
United Kingdom {\textendash} Department of Physics, University of Oxford.
\end{document}